\documentclass[aps,prd,onecolumn,showpacs,showkeys,amsmath,amssymb]{revtex4}
\usepackage{graphicx}
\usepackage{amssymb}
\usepackage{amsmath}
\usepackage{dcolumn}
\usepackage{multirow}

\def\half{\frac{1}{2}}
\def\thalf{\frac{3}{2}}

\def\Sprot#1#2{\Gamma^{#1#2}_{3/2}}

\def\half{\frac{1}{2}}

\def\Sprot#1#2{\Gamma^{#1#2}_{3/2}}

\begin{document}

\title{Bi-Local Baryon Interpolating Fields with Three Flavours}
\author{Hua-Xing Chen$^{1}$}
\email{hxchen@buaa.edu.cn}
\author{V. Dmitra\v sinovi\' c$^2$}
\email{dmitrasin@ipb.ac.rs}
\affiliation{
$^1$ School of Physics and Nuclear Energy Engineering and International Research Center for Nuclei and Particles in the Comos, Beihang University, Beijing 100191, China \\
$^2$ Institute of Physics, Belgrade University, Pregrevica 118, Zemun,
P.O.Box 57, 11080 Beograd,
Serbia}
\begin{abstract}
Fierz identities follow from permutations of quark indices and thus determine
which chiral multiplets of baryon fields are Pauli-allowed, and which are not.
In a previous paper we have investigated the Fierz identities of baryon
fields with two light flavours and found that all bilocal fields that
can be constructed from three quarks are Pauli-allowed. That does not
mean that all possible chiral multiplets exist, however: some chiral
multiplets do not appear among structures with a given spin in the local limit,
say $J = 1/2$.
One such chiral multiplet is the $[(\mathbf{6},\mathbf{3})\oplus (\mathbf{3},\mathbf{6})]$,
which is necessary for a successful chiral mixing  phenomenology.
In the present paper we extend those methods to three light flavors,
i.e. to $SU_F(3)$ symmetry and explicitly construct all three necessary chiral
$SU_L(3) \times SU_R(3)$ multiplets, {\it viz.}
$[(\mathbf{6},\mathbf{3})\oplus (\mathbf{3},\mathbf{6})]$,
$[\mathbf{3},\bar{\mathbf{3}})\oplus(\bar{\mathbf{3}},\mathbf{3})]$ and
$[(\bar{\mathbf{3}},\mathbf{3})\oplus(\mathbf{3},\bar{\mathbf{3}})]$
that are necessary for a phenomenologically successful chiral mixing. We
complete this analysis by considering some bi-local baryon fields that are
sufficient for the construction of the ``missing''  spin $1/2$ baryon interpolating fields.
Bi-local baryon fields have definite total angular momentum only in the local limit.
The physical significance of these results lies in the fact that they
show that there is no need for higher Fock space components, such
as the $q^4 {\bar q}$, in the baryon chiral mixing framework, for the purpose
of fitting the observed axial couplings and magnetic moments: all of the
sufficient ``mirror components'' exist as bi-local fields.
\end{abstract}
\maketitle
\pagenumbering{arabic}
%
%
\label{intro}
\section{Introduction}
%
It is by now fairly well known that the $SU_L(3)\times SU_R(3)$ chiral
symmetry multiplets' mixing successfully describes several basic properties
of $J^P = (\frac12)^+$ baryons, including their Abelian and non-Abelian
axial couplings, and their magnetic moments~\cite{Chen:2008qv,Chen:2009sf,Chen:2010ba,Chen:2011rh}.
For the phenomenological mixing to work one only needs a few (three,
to be precise) out of (five ``naive'' plus five ``mirror'' =) 10 possible
chiral multiplets built from three-quark interpolating fields.
Not all 10 chiral multiplets exist in the local triquark baryon field
limit~\cite{Chen:2008qv,Nagata:2007di,Nagata:2008zzc}, however, due a) to
the fact that some chiral structures
are not associated with all values of spin; and b) to the Pauli exclusion
principle, implemented by way of Fierz identities that
annihilate certain (local) interpolators corresponding to Pauli-forbidden
states. As one relaxes the restriction from strictly local
fields~\cite{Chen:2008qv,Nagata:2007di,Nagata:2008zzc} to
bilocal~\cite{Dmitrasinovic:2011yf},
and finally trilocal fields~\cite{Chen:2012vs}, one may use the additional spatial
degree of freedom to antisymmetrize with, and thus one
finds that some previously Pauli-forbidden two-flavor chiral multiplets
are allowed in the non-local case.
In this manner we found that all chiral structures available for a particular
``value of spin'' are Pauli allowed in bilocal two-flavor baryon sector.
Strictly speaking, rather than the spin it is the Lorentz group representation
that is important here, as for spins higher than 1/2, there is usually more than one
Lorentz group representation (L.g.r.) that corresponds to that particular value of spin,
Ref.~\cite{Y. Ohnuki 88}.

Moreover, some chiral multiplets appear more than once in the non-local
case, whereas in the local limit, they were explicitly shown as
identical by way of Fierz identities. And yet, it is not always possible
to construct all of the ``naive'', or ``mirror'' multiplets from three
non-local quark fields, although generally this can be accomplished using
five-quark, i.e., $q^4 {\bar q}$ fields.
Now, some of the ``missing multiplets'' can be obtained as by-products
of unphysical (spin) degrees of freedom from higher-spin fields'
``projecting out'' procedure. For example, as a by-product of projecting out
the spin-3/2 component from the Rarita-Schwinger [L.g.r.~$(1,1/2)$]
fields, one obtains a spin-1/2 field component with chiral properties
that are ``opposite''/mirror to those of the spin-3/2 component.
This provides the (phenomenologically absolutely necessary) chiral 
$\left[(1,{\frac12}) \oplus ({\frac12},1)\right]$
multiplet in the $J^P = (\frac12)^+$ baryon sector, whereas
the non-local fields provide only the ``mirror'' chiral 
$\left[({\frac12},1) \oplus (1,{\frac12})\right]$multiplet.

With three light flavors there is a bigger variety of both flavor and
chiral multiplets than with two. For this reason one cannot readily
generalize our two-flavor results to three flavors.
So, the question remains if all of the phenomenologically necessary
$SU_L(3) \times SU_R(3)$ chiral multiplets exist
in the three-quark non-local case? In particular the question
of so called ``mirror'' multiplets' existence is important, as they
can be (easily) constructed from (3q + meson) fields, but not necessarily
from three quarks. If such ``mirror'' fields exist only in the (3q + meson)
form, then that would be first indication of a non-exotic ``pentaquark''
Fock component in the nucleon's wave function.
In the present paper we answer that question
for $J^P = (\frac12)^+$ baryons; higher spin objects will not be dealt
with here systematically, except for the explicit purpose of providing
spin-1/2 components.

In a series of previous papers, Refs.~\cite{Nagata:2007di,Nagata:2008zzc,Dmitrasinovic:2011yf,Chen:2012vs},
we have investigated the Fierz identities and chiral $SU_L(2)\times SU_R(2)$
transformation properties of bilocal baryon fields with two light
flavours. In the present paper we extend those methods and results
to three light flavors, i.e. to $SU(3)_F$ symmetry.

We note here that this extension to three flavors introduces only a mathematical change to
the analogous two-flavor analysis, Ref. \cite{Chen:2011rh,Chen:2012ex}:
the fact that the $SU(3)_F$ symmetry is explicitly broken does not play a role here,
because the quark mass difference does not enter into considerations of the permutation symmetry.
Rather, it is the very existence of the third flavor that makes the difference.
Needless to say, the most remarkable consequences are in the flavor-singlet
channel that does not exist with two flavors. Another place where the difference
between two and three flavors is pronounced are the flavor octet chiral multiplets
$\left[({\bf 8},{\bf 1}) \oplus ({\bf 1}, {\bf 8})\right]$, and
$\left[({\bf 3},\overline{{\bf 3}}) \oplus (\overline{{\bf 3}}, {\bf
3})\right]$, both of which are ``reduced to'' the two-flavor chiral multiplet
$\left[({\frac12},{0}) \oplus ({0}, {\frac12})\right]$.

Whereas the $SU(3)$ algebra is considerably more complicated than the $SU(2)$ one,
the physical results are largely determined by the overall
permutation symmetry properties (i.e. the Fierz identities) of the
baryon operators, which, in turn, are determined by the chiral
$SU_L(3)\times SU_R(3)$, or $SU_L(2)\times SU_R(2)$ multiplets.
As the $SU_L(3)\times SU_R(3)$ multiplets contain (several
smaller) $SU_L(2)\times SU_R(2)$ multiplets within them, that have already
been examined in Refs.~\cite{Dmitrasinovic:2011yf,Chen:2012vs} it should come as
no surprise that the $SU_L(3)\times SU_R(3)$ ``completions'' of chiral
$SU_L(2)\times SU_R(2)$ multiplets exist as well.
Indeed, one may adopt a chiral multiplet nomenclature based on the
Young diagrams/tableaux, see Table \ref{tab:summary}, rather
than the actual dimensionality of the multiplet,
that shows the full analogy of chiral multiplets with
different flavor numbers.
There is (only) one exception to this $SU(3)$ completion ``rule'':
the flavor-singlet $[(\mathbf{1},\mathbf{1})]$, $\Lambda$ hyperon,
that is antisymmetric in flavor space, and does not exist
with two flavors. It can either belong to a chiral
$\left[({\bf 3},\overline{{\bf 3}}) \oplus (\overline{{\bf 3}}, {\bf
3})\right]$ multiplet, or to a chiral singlet.

The primary question is then: which chiral multiplets do
these (``new'') bi-local operators belong to? We investigate all
cases and classify the bi-local three-flavor baryon interpolators
according to their chiral transformations. Before doing that, we would like to note that the bi-local or tri-local fields
have components overlapping with more than one orbital angular momentum $L$ states. To project out definite-J components from these fields,
one needs to specify the three-body dynamics. For example, if one wishes to use such fields on the lattice, one can use the
Euclidean space version, and the corresponding spin projection methods, such as that in Ref.~\cite{Basak:2005ir}.
However, these operators have definite total angular momentum only in the limit of local fields, and so we shall assume
our non-local fields have spins $J=1/2$ or $J=3/2$ in the following analysis.

We find three new spin
$1/2$ chiral multiplets that do not exist in the local-operator
limit: one $[(\mathbf{6},\mathbf{3})\oplus (\mathbf{3},\mathbf{6})]$, one
$[(\mathbf{10},\mathbf{1})\oplus(\mathbf{1},\mathbf{10})]$ and one
$[(\mathbf{1},\mathbf{1})]$, and several other multiplets that used to
be related (``identical'') by Fierz identities to others, that are
independent in the non-local case.
The chiral transformations do not depend on the (non-)locality of
the operator, but the Fierz identities do. For this reason we
concentrate only on the latter in this paper - the
$SU_L(3) \times SU_R(3)$ chiral transformations have been
worked out in some detail in Ref. \cite{Chen:2008qv} and are
briefly reviewed in Appendix~\ref{sec:chiral_baryon}.
The physical significance of our results is that they show an absence of need for
$q^4 {\bar q}$ components when fitting the observed axial couplings and magnetic
moments in the chiral
mixing framework: all of the ``mirror components'' exist as bi-local fields.

This paper consists of four sections and is organized as follows.
After the (present) Introduction in Sect.~\ref{sec:straight_fields},
we firstly define all possible ``straightforward bi-local extensions''
of local baryon operators. There we classify the baryon operators
according to the representations of the Lorentz and the flavor groups,
{\it viz.} the Dirac, the Rarita-Schwinger (RS) and the antisymmetric
tensor (a.s.t.) Bargmann-Wigner (BW) fields. Then in Sect.~\ref{sec:non_straight_fields},
we define the ``non-straightforward bi-local extensions''
of local baryon operators, such as the derivative-contracted
RS and a.s. tensor fields, that appear as by-products of
spin $3/2$ projecting out.
The final section \ref{sec:summary} is a summary and an outlook
to possible future extensions and applications.
In Appendix~\ref{sec:chiral_baryon}, we define the
Abelian and non-Abelian chiral transformations of the baryon
operators as functions of the quarks' chiral transformation
parameters.

\section{Straightforward Three-Flavor Bi-local Three-Quark Fields}
\label{sec:straight_fields}

Three-quark baryon interpolating fields in QCD have well-defined
$SU_L(3) \times SU_R(3)$ and $U_A(1)$ chiral transformation
properties, see Table \ref{tab:summary},
\begin{eqnarray} \left[({\bf 3}, {\bf 1}) \oplus ({\bf 1},{\bf 3})
\right]^3 \sim
\left[({\bf 1},{\bf 1})\right] \oplus
\left[({\bf 8},{\bf 1}) \oplus ({\bf 1}, {\bf 8})\right] \oplus
\left[({\bf 10},{\bf 1}) \oplus ({\bf 1}, {\bf 10})\right]
\oplus
\left[({\bf 6},{\bf 3}) \oplus ({\bf 3},{\bf 6}) \right] \oplus
\left[({\bf 3},\overline{{\bf 3}}) \oplus (\overline{{\bf 3}}, {\bf
3})\right]\, ,
\end{eqnarray}
{\it viz.} $[({\bf 6},{\bf 3})\oplus({\bf 3},{\bf 6})]$,
$[({\bf 3},\overline{{\bf 3}}) \oplus (\overline{{\bf 3}}, {\bf
3})]$, $[({\bf 1},{\bf 1})]$,
$[({\bf 8},{\bf 1}) \oplus ({\bf 1}, {\bf 8})]$,
$[({\bf 10},{\bf 1}) \oplus ({\bf 1}, {\bf 10})]$, and their
``mirror'' images, Ref.~\cite{Chen:2008qv}. It has been shown
(phenomenologically) in Ref.~\cite{Chen:2009sf} that mixing of the
$[({\bf 6},{\bf 3})\oplus({\bf 3},{\bf 6})]$ chiral multiplet with
one ordinary (``naive'') $[({\bf 3},\overline{{\bf 3}}) \oplus
(\overline{{\bf 3}}, {\bf 3})]$ and one ``mirror'' field
$[(\overline{{\bf 3}}, {\bf 3}) \oplus ({\bf 3},\overline{{\bf 3}})]$
multiplet can be used to fit the values of the isovector ($g_A^{(3)}$) and the
flavor-singlet (isoscalar) axial coupling ($g_A^{(0)}$) of the
nucleon and then predict the axial $F$ and $D$ coefficients, or {\it
vice versa}, in reasonable agreement with experiment.
\begin{table}[tbh]
\renewcommand{\arraystretch}{1.2}
\begin{center}
\caption{Structure of all three-quark baryon fields in the local
limit, together with their Lorentz group representation, spin, Young diagram,
chiral $SU(2)$ and $SU(3)$ representations, axial $U(1)_A$ charge $g^0_A$ and
their Fierz transformation equivalent fields, or vanishing for Pauli-forbidden fields.}
\begin{tabular}{c | c | c | c | c | c | c | c }
\hline \hline
Lorentz & Spin &
$\begin{array}{c} \mbox{Young diagram} \\ \mbox{for Chiral rep.} \end{array}$ &
Chiral $SU(2)$ & Chiral $SU(3)$ & $g^0_A$
& Fields & $\begin{array}{c} \mbox{Fierz}\& \\ \mbox{Local Lim.} \end{array}$ \\
\hline \hline
\multirow{7}{*}{$\begin{array}{c}(\frac12,0)\oplus\\(0,\frac12)\end{array}$} & \multirow{7}{*}{$1/2$} &
$([111],-)\oplus(-,[111])$ & \multirow{6}{*}{$(\frac12,0)\oplus(0,\frac12)$} &
$(\mathbf{1},\mathbf{1})$ & $~~3$ & $\Lambda_1 + \Lambda_2$ & 0 \\
\cline{3-3} \cline{4-4}
\cline{5-8} & & \multirow{2}{*}{$([21],-)\oplus(-,[21])$} & &
\multirow{2}{*}{$(\mathbf{8},\mathbf{1})\oplus(\mathbf{1},\mathbf{8})$} &
\multirow{2}{*}{$~~3$} & $N_1 + N_2$ & \multirow{2}{*}{$N_1 + N_2$}
\\ \cline{7-7} & & & & & & $M_5$ &
\\ \cline{3-3} \cline{5-8} & & \multirow{2}{*}{$([1],[11])\oplus([11],[1])$} & &
\multirow{2}{*}{$(\mathbf{3},\bar{\mathbf{3}})\oplus(\bar{\mathbf{3}},\mathbf{3})$} &
\multirow{2}{*}{$-1$} & $(\Lambda_1 - \Lambda_2, N_1 - N_2)$ & \multirow{2}{*}{$(\Lambda_1, N_1 - N_2)$}
\\ \cline{7-7} & & & & & & $(\Lambda_3, N_3 - M_4)$ &
\\ \cline{3-8} & & $([1],[2])\oplus([2],[1])$ & $(\frac12,1)\oplus(1,\frac12)$ &
$(\mathbf{3},\mathbf{6})\oplus(\mathbf{6},\mathbf{3})$ & $-1$ &
$(N_3 + \frac13 M_4, \Delta_4)$ & 0
\\ \cline{3-8} & & $([3],-)\oplus(-,[3])$ & $(\frac32,0)\oplus(0,\frac32)$ &
$(\mathbf{10},\mathbf{1})\oplus(\mathbf{1},\mathbf{10})$ & $~~3$ & $\Delta_5$ & 0
\\ \hline \hline
\multirow{3}{*}{$\begin{array}{c}(\frac12,1)\oplus\\(1,\frac12)\end{array}$} &
\multirow{3}{*}{$\begin{array}{c}1/2\\\&\\3/2\end{array}$} & $([11],[1])\oplus([1],[11])$ &
$(0,\frac12)\oplus(\frac12,0)$
& $(\bar{\mathbf{3}},\mathbf{3})\oplus(\mathbf{3},\bar{\mathbf{3}})$ & $~~1$ & $(\Lambda_3^\mu, N_3^\mu - M_4^\mu)$ & 0
\\
\cline{3-8} & & \multirow{2}{*}{$([2],[1])\oplus([1],[2])$} &
\multirow{2}{*}{$(1,\frac12)\oplus(\frac12,1)$} & \multirow{2}{*}{$(\mathbf{6},\mathbf{3})\oplus(\mathbf{3},\mathbf{6})$}
& \multirow{2}{*}{$~~1$} & $(N_3^\mu + \frac13 M_4^\mu, \Delta_4^\mu)$ & \multirow{2}{*}{$(N_3^\mu, \Delta_4^\mu)$}
\\
\cline{7-7} & & & & & & $(M_5^\mu, \Delta_5^\mu)$ &
\\
\hline \hline
\multirow{2}{*}{$\begin{array}{c}(\frac32,0)\oplus\\(0,\frac12)\end{array}$} &
\multirow{2}{*}{$3/2$} & $([21],-)\oplus(-,[21])$ & $(\frac12,0)\oplus(0,\frac12)$ &
$(\mathbf{8},\mathbf{1})\oplus(\mathbf{1},\mathbf{8})$ & $~~3$ & $M_5^{\mu\nu}$ & 0
\\ \cline{3-8} & & $([3],-)\oplus(-,[3])$ & $(\frac32,0)\oplus(0,\frac32)$ &
$(\mathbf{10},\mathbf{1})\oplus(\mathbf{1},\mathbf{10})$ & $~~3$ & $\Delta_5^{\mu\nu}$ & $\Delta_5^{\mu\nu}$
\\
\hline \hline
\end{tabular}
\label{tab:summary}
\end{center}
\renewcommand{\arraystretch}{1}
\end{table}
Moreover, this mixing can be reproduced by a chirally symmetric interaction
Lagrangian with observed baryon masses used as the input for unknown
coupling constants, Ref.~\cite{Chen:2010ba}, and the anomalous magnetic
moments of baryons can be introduced in according with chiral symmetry
and experimental observations, Ref.~\cite{Chen:2011rh}.
For this reason it is vital that all three of these chiral multiplets
are not forbidden by the Pauli principle in the three-quark interpolators.
Yet, the original analysis of local three-quark fields, Ref.~\cite{Chen:2008qv},
allowed only one out of three: the  (``naive'') $[({\bf 3},\overline{{\bf 3}}) \oplus
(\overline{{\bf 3}}, {\bf 3})]$. In the following we shall explicitly
construct the other two interpolators. For that purpose we shall need
both the straightforward and the not-so-straightforward extensions
of local fields, as the straightforward method yields only the ``mirror''
field $[(\overline{{\bf 3}}, {\bf 3}) \oplus ({\bf 3},\overline{{\bf 3}})]$,
whereas the $[({\bf 6},{\bf 3})\oplus({\bf 3},{\bf 6})]$ chiral multiplet
appears only as a remnant of the spin-projection procedure in Rarita-Schwinger
fields.

Before doing that, we would like to note that the bi-local or tri-local fields contain
in general (infinitely many) components overlapping with more than one orbital angular
momentum $L$ state.
Consequently, these operators have definite total angular momentum $J$ only
in the limit of local fields, though individual $J$ components might be
extracted by a suitable spin-projection. Such a spin projection technique
has been devised for three-quark fields on a Euclidean lattice space-time, Ref.
\cite{Basak:2005ir},
though in a continuum Minkowski space-time, one is better suited by projecting
out good-$J$ states in matrix elements, e.g. using the Jacob-Wick formalism,
rather than in operators themselves.
In order to project out the good-$J$ operators and thus address this ``theoretical
uncertainty'', one has to specify the three-body dynamics explicitly, which
is well beyond the scope of this paper..

At any rate, such a total angular momentum projection
would not change the Dirac structure of the composite fields, and
their chiral properties would remain unchanged, as well. Moreover, the existence
of the two lowest values ($J=1/2$ or $J=3/2$) of the total angular
momentum $J$ components in our non-local fields is beyond doubt, anyway.

%
\subsection{Dirac fields}
\label{subsec:Dirac}
%

In this section we investigate independent baryon fields for each
Lorentz group representation which is formed by three quarks. The
Clebsch-Gordan series for the irreducible decomposition of the
direct product of three $(\frac12, 0) \oplus (0,\frac12)$
representations of the Lorentz group (the three quark Dirac fields)
is
\begin{eqnarray} \left((\frac12, 0) \oplus (0,\frac12)\right)^3 \sim
\left( (\frac12, 0) \oplus (0,\frac12) \right) \oplus
\left((1,\frac12) \oplus (\frac12,1) \right)
\oplus \left( (\frac32, 0) \oplus (0,\frac32) \right)\, ,
\end{eqnarray}
where we have ignored the different multiplicities of the
representations on the right-hand side. Three Lorentz group representations
$\left( (\frac12, 0) \oplus (0,\frac12) \right)$, $\left((1,
\frac12) \oplus (\frac12,1) \right)$, $\left( (\frac32, 0) \oplus
(0,\frac32) \right)$ describe the Dirac spinor field, the
Rarita-Schwinger's vector-spinor field and the
antisymmetric-tensor-spinor field, respectively. In order to
establish independent fields we employ the Fierz transformations for
the color, flavor, and Lorentz (spin) degrees of freedom, which is
essentially equivalent to the Pauli principle for three quarks. Here
we demonstrate the essential idea for the simplest case of the Dirac
spinor, $ (\frac12, 0) \oplus (0,\frac12) $.

It is convenient to introduce a ``tilde-transposed'' quark field
$\tilde{q}$ as follows
\begin{eqnarray}
\tilde{q}=q^T C\gamma_5 \, ,
\end{eqnarray}
where $C = i\gamma_2\gamma_0$ is the Dirac field charge conjugation
operator.

\subsubsection{Flavor singlet baryon}

Let us start by writing down five tri-local baryon fields that contain a ``diquark''
operator, formed by one of five sets of (products of) Dirac matrices,
$1, \gamma_5, \gamma_\mu, \gamma_\mu \gamma_5$ and $\sigma_{\mu \nu}$,
\begin{eqnarray}
\begin{array}{l}
\Lambda_1(x,y,z) = \epsilon_{abc} \epsilon^{ABC} (\tilde{q}_A^a(x) q_B^b(y)) q_C^c(z) \, ,
\\ \Lambda_2(x,y,z) = \epsilon_{abc} \epsilon^{ABC} (\tilde{q}_A^a(x) \gamma_5 q_B^b(y)) \gamma_5 q_C^c(z) \, ,
\\ \Lambda_3(x,y,z) = \epsilon_{abc} \epsilon^{ABC} (\tilde{q}_A^a(x) \gamma_\mu q_B^b(y)) \gamma^\mu q_C^c(z) \, ,
\\ \Lambda_4(x,y,z) = \epsilon_{abc} \epsilon^{ABC} (\tilde{q}_A^a(x) \gamma_\mu \gamma_5 q_B^b(y)) \gamma^\mu \gamma_5 q_C^c(z) \, ,
\\ \Lambda_5(x,y,z) = \epsilon_{abc} \epsilon^{ABC} (\tilde{q}_A^a(x) \sigma_{\mu\nu} q_B^b(y)) \sigma_{\mu\nu} q_C^c(z)
\, .
\end{array}
\label{e:Lambda12}
\end{eqnarray}
Here and in the following we use the notation and conventions of Sect. II in Ref. \cite{Chen:2008qv},
where the capital roman letter indices e.g. $A,B,C=1,2,3$ denote the $SU(3)$ flavor degrees of
freedom of a quark, and $\epsilon^{ABC}$ is the (Levy-Civita) totally antisymmetric tensor.
The antisymmetric tensor in color space $\epsilon_{abc}$, ensures the baryons' being color singlets.
Our results are not affected by taking non-local baryon operators with path-ordered phase factors
\begin{eqnarray}
B(x_1 , x_2 , x_3)   &\sim& \epsilon_{abc} (\tilde{q}_{a'}(x_1) q_{b'}(x_2)) q_{b'}(x_3)
\times \left[ P {\rm exp}\left(i g\int_{x_1}^{z} ~A_{\mu}(y_1)dy_1^{\mu}
\right) \right]_{a'a}
\nonumber \\
&& \times \left[ P {\rm exp}\left(i g\int_{x_2}^{z} ~A_{\mu}(y_2)dy_2^{\mu}
\right) \right]_{b'b}
\times \left[ P {\rm exp}\left(i g\int_{x_3}^{z} ~A_{\mu}(y_3)dy_3^{\mu}
\right) \right]_{c'c}
\,
\label{e:colorSU3phases}
\end{eqnarray}
that ensure local SU(3) color invariance, cf. Ref. \cite{Christos:1986us},
instead of the straightforward ones, such as those in Eq. (\ref{e:Lambda12}).
As these factors are always assumed to be present, we shall omit them from now on, but
we note that they give an extra minus sign when performing a color SU(3) Fierz transformation.

Due to the non-locality of these operators, the Pauli principle does not
forbid any one of these {\it a priori}.
For each one of the five tri-local operators $\Lambda_i(x,y,z)$ in Eq. (\ref{e:Lambda12}), there
are three possible fields with bi-local (functions of two position four-vectors $x$ and $y$)
operators:
\begin{eqnarray}
\Lambda_i(x,x,y) \, , \, \Lambda_i(x,y,x) \, , \, \Lambda_i(y,x,x) \, .
\end{eqnarray}
The latter two sets can be related to each other by simply
interchanging the positions of the first and second quark fields, for example,
\begin{eqnarray}\label{eq:relation}
q^{aT}_A(x) \gamma_5 q^b_B(y) = - q^{bT}_B(y) \gamma_5 q^a_A(x) \, .
\end{eqnarray}
The last two are also related to the first set through the Fierz transformation:
\begin{eqnarray}
\Lambda_j(x,y,x) &=& T^{S1}_{ij} \Lambda_i(x,x,y) \, ,
\end{eqnarray}
where the transition matrix ${\bf T}^{S1}$ is
\begin{eqnarray}
{\bf T}^{S1} = {\frac14} \left (
\begin{array}{ccccc}
-1 & -1 & -1 & -1 & {\frac12}
\\ -1 & -1 & 1 & 1 & {\frac12}
\\ -4 & 4 & 2 & -2 & 0
\\ 4 & -4 & 2 & -2 & 0
\\ -12 & -12 & 0 & 0 & -2
\end{array} \right ) \, .
\label{e:TS1}
\end{eqnarray}
The Pauli principle does eliminate some local diquarks, however, and one quickly finds that
\begin{eqnarray}
\Lambda_4(x,x,y) = \Lambda_5(x,x,y) = 0 \, .
\end{eqnarray}
Therefore, only three of the original 15 operators are independent. They are:
$\Lambda_1(x,x,y)$, $\Lambda_2(x,x,y)$ and $\Lambda_3(x,x,y)$.

\subsubsection{The flavour decuplet baryons}

There are also five decuplet baryon fields formed from five different combinations
of $\gamma$-matrices:
\begin{eqnarray}
\begin{array}{l}
\Delta^P_1 = S_P^{ABC} (\tilde{q}_A q_B) q_C \, ,
\\ \Delta^P_2 = S_P^{ABC} (\tilde{q}_A \gamma_5 q_B) \gamma_5 q_C \, ,
\\ \Delta^P_3 = S_P^{ABC} (\tilde{q}_A \gamma_\mu q_B) \gamma^\mu q_C \, ,
\\ \Delta^P_4 = S_P^{ABC} (\tilde{q}_A \gamma_\mu \gamma_5 q_B) \gamma^\mu \gamma_5 q_C \, ,
\\ \Delta^P_5 = S_P^{ABC} (\tilde{q}_A \sigma_{\mu\nu} q_B) \sigma_{\mu\nu} q_C \, .
\end{array}
\end{eqnarray}
Here $S_P^{ABC}$ is the totally symmetric $SU(3)$ tensor with components
listed in Table~\ref{tab:SP_ABC}. Index $P=1,...,10$, denotes the $SU(3)$ flavor label
of a decuplet state.
\begin{table}[hbt]
\begin{center}
\caption{Non-zero components of $S_P^{ABC}$}
\begin{tabular}{c | c c c c c c c c c c}
\hline \hline $P$&1&2&3&4&5&6&7&8&9&10
\\ \hline $ABC$ & 111 & 112 & 122 & 222 & 113 & 123 & 223 & 133 & 233
& 333
\\ \hline Baryons & $\Delta^{++}$ & $\Delta^{+}$ & $\Delta^{0}$ & $\Delta^{-}$ &
$\Sigma^{*+}$ & $\Sigma^{*0}$ & $\Sigma^{*-}$ & $\Xi^{*0}$ &
$\Xi^{*-}$ & $\Omega^{-}$
\\ \hline Normalization & 1 & $\frac{1}{\sqrt3}$ & $\frac{1}{\sqrt3}$ & 1 &
$\frac{1}{\sqrt3}$ & $\sqrt6$ & $\frac{1}{\sqrt3}$ & $\frac{1}{\sqrt3}$ &
$\frac{1}{\sqrt3}$ & 1
\\ \hline
\end{tabular}
\label{tab:SP_ABC}
\end{center}
\end{table}
Here also we have three sets of bi-local fields, that are related to each
other by Fierz identities:
\begin{eqnarray}
\nonumber \Delta^P_i(y,x,x) &\leftrightarrow& \Delta^P_j(x,y,x) \, , \\
\nonumber \Delta^P_j(x,y,x) &=& T^{D1}_{ij} \Delta^P_i(x,x,y) \, ,
\end{eqnarray}
where the flavor-decuplet matrix ${\bf T}^{D1}$ is identical to the
flavor-singlet matrix ${\bf T}^{S1}$ given in Eq. (\ref{e:TS1}),
\begin{eqnarray}
{\bf T}^{D1} = {\bf T}^{S1} \, .
\end{eqnarray}
Due to the Pauli principle, we find that
\begin{eqnarray}
\Delta^P_1(x,x,y) = \Delta^P_2(x,x,y) = \Delta^P_3(x,x,y) = 0 \, .
\end{eqnarray}
Therefore, only two of the original 15 bi-local $\Delta$ operators are
independent. They are: $\Delta^P_4(x,x,y)$ and $\Delta^P_5(x,x,y)$.

\subsubsection{The flavor octet baryon fields}
\label{subsub:octet}

We start once again with five tri-local fields
\begin{eqnarray}
\begin{array}{l}
N_1^N = \epsilon^{ABD} \lambda_{DC}^N (\tilde{q}_A q_B) q_C \, ,
\\ N_2^N = \epsilon^{ABD} \lambda_{DC}^N (\tilde{q}_A \gamma_5 q_B) \gamma_5 q_C \, ,
\\ N_3^N = \epsilon^{ABD} \lambda_{DC}^N (\tilde{q}_A \gamma_\mu q_B) \gamma^\mu q_C \, ,
\\ N_4^N = \epsilon^{ABD} \lambda_{DC}^N (\tilde{q}_A \gamma_\mu \gamma_5 q_B) \gamma^\mu \gamma_5 q_C \, ,
\\ N_5^N = \epsilon^{ABD} \lambda_{DC}^N (\tilde{q}_A \sigma_{\mu\nu} q_B) \sigma_{\mu\nu} q_C \, .
\end{array}
\end{eqnarray}
The index $N=1,...,8$, labels the flavor $SU(3)$ states in an octet.
Here $\lambda_{DC}^N$ is the $D$-th column, $C$-th row component of the $N$-th Gell-Mann matrix.
There are, however, two other kinds of baryon octet fields with the flavor $SU(3)$ structures
$\epsilon^{BCD} \lambda_{DA}^N$ and $\epsilon^{CAD} \lambda_{DB}^N$:
\begin{eqnarray}
\begin{array}{l}
N_6^N = \epsilon^{BCD} \lambda_{DA}^N (\tilde{q}_A q_B) q_C \, ,
\\ N_7^N = \epsilon^{BCD} \lambda_{DA}^N (\tilde{q}_A \gamma_5 q_B) \gamma_5 q_C \, ,
\\ N_8^N = \epsilon^{BCD} \lambda_{DA}^N (\tilde{q}_A \gamma_\mu q_B) \gamma^\mu q_C \, ,
\\ N_9^N = \epsilon^{BCD} \lambda_{DA}^N (\tilde{q}_A \gamma_\mu \gamma_5 q_B) \gamma^\mu \gamma_5 q_C \, ,
\\ N_{10}^N = \epsilon^{BCD} \lambda_{DA}^N (\tilde{q}_A \sigma_{\mu\nu} q_B) \sigma_{\mu\nu} q_C \, .
\end{array}
\begin{array}{l}
N_{11}^N = \epsilon^{CAD} \lambda_{DB}^N (\tilde{q}_A q_B)  q_C \, ,
\\ N_{12}^N = \epsilon^{CAD} \lambda_{DB}^N (\tilde{q}_A \gamma_5 q_B) \gamma_5 q_C \, ,
\\ N_{13}^N = \epsilon^{CAD} \lambda_{DB}^N (\tilde{q}_A \gamma_\mu q_B) \gamma^\mu q_C \, ,
\\ N_{14}^N = \epsilon^{CAD} \lambda_{DB}^N (\tilde{q}_A \gamma_\mu \gamma_5 q_B) \gamma^\mu \gamma_5 q_C \, ,
\\ N_{15}^N = \epsilon^{CAD} \lambda_{DB}^N (\tilde{q}_A \sigma_{\mu\nu} q_B) \sigma_{\mu\nu} q_C \, .
\end{array}
\end{eqnarray}
We have to consider all three sets of bi-local fields; they are related 
through the Fierz relation:
\begin{eqnarray}
\nonumber N^N_i(y,x,x) &\leftrightarrow& N^N_i(x,y,x) \, ,
\\ \nonumber N^N_i(x,y,x) &=& T^{O1}_{ij} N^N_i(x,x,y) \, ,
\end{eqnarray}
where the transition matrix ${\bf T}^{O1}$ is obtained from the Fierz transformation
\begin{eqnarray}
{\bf T}^{O1} = \frac{1}{4} \left ( \begin{array}{ccccc|ccccc|ccccc} 0
& 0 & 0 & 0 & 0 & -1 & -1 & -1 & -1 & \frac{1}{2} & 0 & 0 & 0 & 0 & 0
\\ 0 & 0 & 0 & 0 & 0 & -1 & -1 & 1 & 1 & \frac{1}{2} & 0 & 0 & 0 & 0 & 0
\\ 0 & 0 & 0 & 0 & 0 & -4 & 4 & 2 & -2 & 0 & 0 & 0 & 0 & 0 & 0
\\ 0 & 0 & 0 & 0 & 0 & 4 & -4 & 2 & -2 & 0 & 0 & 0 & 0 & 0 & 0
\\ 0 & 0 & 0 & 0 & 0 & -12 & -12 & 0 & 0 & -2 & 0 & 0 & 0 & 0 & 0
\\ \hline 0
& 0 & 0 & 0 & 0 & 0 & 0 & 0 & 0 & 0 & -1 & -1 & -1 & -1 & \frac{1}{2}
\\ 0 & 0 & 0 & 0 & 0 & 0 & 0 & 0 & 0 & 0 & -1 & -1 & 1 & 1 & \frac{1}{2}
\\ 0 & 0 & 0 & 0 & 0 & 0 & 0 & 0 & 0 & 0 & -4 & 4 & 2 & -2 & 0
\\ 0 & 0 & 0 & 0 & 0 & 0 & 0 & 0 & 0 & 0 & 4 & -4 & 2 & -2 & 0
\\ 0 & 0 & 0 & 0 & 0 & 0 & 0 & 0 & 0 & 0 & -12 & -12 & 0 & 0 & -2
\\ \hline  -1 & -1 & -1 & -1 & \frac{1}{2} & 0
& 0 & 0 & 0 & 0 & 0 & 0 & 0 & 0 & 0
\\ -1 & -1 & 1 & 1 & \frac{1}{2} & 0 & 0 & 0 & 0 & 0 & 0 & 0 & 0 & 0 & 0
\\ -4 & 4 & 2 & -2 & 0 & 0 & 0 & 0 & 0 & 0 & 0 & 0 & 0 & 0 & 0
\\ 4 & -4 & 2 & -2 & 0 & 0 & 0 & 0 & 0 & 0 & 0 & 0 & 0 & 0 & 0
\\ -12 & -12 & 0 & 0 & -2 & 0 & 0 & 0 & 0 & 0 & 0 & 0 & 0 & 0 & 0
\end{array} \right ) \, .
\end{eqnarray}
Together with the Jacobi-type identity
\begin{eqnarray}
\epsilon^{ABD} \lambda_{DC}^N + \epsilon^{BCD} \lambda_{DA}^N + \epsilon^{CAD}
\lambda_{DB}^N = 0 \, , \label{eq:Jacobi}
\end{eqnarray}
and the Pauli principle, we obtain (only) five of the original
15 operators that are independent. Here we choose them as
$N_1(x,x,y)$, $N_2(x,x,y)$, $N_3(x,x,y)$ and
\begin{eqnarray}
M_4(x,x,y) &=& N_9(x,x,y) - N_{14}(x,x,y) \, ,
\\ M_5(x,x,y) &=& N_{10}(x,x,y) - N_{15}(x,x,y) \, .
\end{eqnarray}
Other octet baryons can be related to these five; here we only
show the equations for $N_6(x,x,y)$, $N_7(x,x,y)$ and $N_8(x,x,y)$:
\begin{eqnarray}
N_6(x,x,y) &=& - \frac{1}{2} N_1(x,x,y) \, ,
\\ N_7(x,x,y) &=& - \frac{1}{2} N_2(x,x,y) \, ,
\\ N_8(x,x,y) &=& - \frac{1}{2} N_3(x,x,y) \, .
\end{eqnarray}

%
\subsection{Rarita-Schwinger Fields}
\label{subsec:RS}
%

\subsubsection{Flavor singlet baryon}

We start by writing down three trilocal baryon fields
\begin{eqnarray}
\begin{array}{l}
\Lambda_{3\mu} = \epsilon^{ABC} (\tilde{q}_A \gamma_\nu q_B)
\Gamma^{\mu\nu}_{3/2} \gamma_5 q_C \, ,
\\ \Lambda_{4\mu} = \epsilon^{ABC} (\tilde{q}_A \gamma_\nu \gamma_5 q_B)
\Gamma^{\mu\nu}_{3/2} q_C  \, ,
\\ \Lambda_{5\mu} = \epsilon^{ABC} (\tilde{q}_A \sigma_{\alpha\beta} q_B)
\Gamma^{\mu\alpha}_{3/2}
\gamma^\beta \gamma_5 q_C \, ,
\end{array}
\end{eqnarray}
that contain diquarks formed from three sets of Dirac matrices, $\gamma_\mu,
\gamma_\mu \gamma_5$ and $\sigma_{\mu \nu}$.
Here $\Sprot{\mu}{\nu}$ is the projection operator for the Rarita-Schwinger
fields:
\begin{eqnarray}
\Sprot{\mu}{\nu}=g^{\mu\nu}-\frac14 \gamma^\mu\gamma^\nu \, .
\end{eqnarray}
Here, again, we have three sets of bi-local fields that are related to each other
through the Fierz transformation:
\begin{eqnarray}
\nonumber \Lambda_{i\mu}(y,x,x) &\leftrightarrow& \Lambda_{j\mu}(x,y,x) \, ,
\\ \nonumber \Lambda_{i\mu}(x,y,x) &=& T^{S2}_{ij} \Lambda_{j\mu}(x,x,y) \, ,
\end{eqnarray}
where the transition matrix ${\bf T}^{S2}$ is
\begin{eqnarray}
{\bf T}^{S2} = \frac{1}{2} \left ( \begin{array}{ccc} -1 & 1 & 1
\\ -1 & 1 & -1
\\ -2 & -2 & 0
\end{array} \right ) \,
\label{e:TS2}.
\end{eqnarray}
Due to the Pauli principle, we find vanishing of two fields
\begin{eqnarray}
\Lambda_{4\mu}(x,x,y) = \Lambda_{5\mu}(x,x,y) = 0 \, .
\end{eqnarray}
leaving the $\Lambda_{3\mu}(x,x,y)$ as the only non-vanishing bi-local
$\Lambda_{\mu}(x,x,y)$ field. Therefore, only one of the original nine
operators is independent.

\subsubsection{The flavour decuplet baryons}

Let us start by writing down three baryon fields which contain a
diquark formed by three sets of Dirac matrices,
$\gamma_\mu, \gamma_\mu \gamma_5$ and $\sigma_{\mu \nu}$,
\begin{eqnarray}
\begin{array}{l}
\Delta^P_{3\mu} = S_P^{ABC} (\tilde{q}_A \gamma_\nu q_B) \Gamma^{\mu\nu}_{3/2} \gamma_5 q_C \, ,
\\ \Delta^P_{4\mu} = S_P^{ABC} (\tilde{q}_A \gamma_\nu \gamma_5 q_B) \Gamma^{\mu\nu}_{3/2} q_C  \, ,
\\ \Delta^P_{5\mu} = S_P^{ABC} (\tilde{q}_A \sigma_{\alpha\beta} q_B) \Gamma^{\mu\alpha}_{3/2}
\gamma^\beta \gamma_5 q_C \, .
\end{array}
\end{eqnarray}
We have also three sets of bi-local fields that are related 
through the Fierz transformation:
\begin{eqnarray}
\nonumber \Delta^P_{i\mu}(y,x,x) &\leftrightarrow& \Delta^P_{j\mu}(x,y,x) \, ,
\\ \nonumber \Delta^P_{i\mu}(x,y,x) &=& T^{D2}_{ij} \Delta^P_{j\mu}(x,x,y) \, ,
\end{eqnarray}
where the flavor-decuplet Fierz matrix ${\bf T}^{D2}$ is identical to the flavor-singlet
Fierz matrix, Eq. (\ref{e:TS2})
\begin{eqnarray}
{\bf T}^{D2} = {\bf T}^{S2} \, .
\end{eqnarray}
Due to the Pauli principle, we immediately find
\begin{eqnarray}
\Delta^P_{3\mu}(x,x,y) = 0 \, .
\end{eqnarray}
Therefore, only two ($\Delta^P_{4\mu}(x,x,y)$ and $\Delta^P_{5\mu}(x,x,y)$)
of the original nine operators are independent.

\subsubsection{The flavor octet baryon fields}
\label{subsub:octetRS}

Again, we start by writing down three trilocal baryon fields
\begin{eqnarray}
\begin{array}{l}
N_{3\mu}^N = \epsilon^{ABD} \lambda_{DC}^N (\tilde{q}_A \gamma_\nu q_B) \Gamma^{\mu\nu}_{3/2} \gamma_5 q_C \, ,
\\ N_{4\mu}^N = \epsilon^{ABD} \lambda_{DC}^N (\tilde{q}_A \gamma_\nu \gamma_5 q_B) \Gamma^{\mu\nu}_{3/2} q_C \, ,
\\ N_{5\mu}^N = \epsilon^{ABD} \lambda_{DC}^N (\tilde{q}_A \sigma_{\alpha\beta} q_B)
\Gamma^{\mu\alpha}_{3/2} \gamma^\beta \gamma_5 q_C \, ,
\end{array}
\end{eqnarray}
that contain a diquark formed with one three sets of Dirac matrices,
$\gamma_\mu, \gamma_\mu \gamma_5$ and $\sigma_{\mu \nu}$.
There are, however, two other kinds of octet baryons with the flavor
structures $\epsilon^{BCD} \lambda_{DA}^N$ and $\epsilon^{CAD} \lambda_{DB}^N$:
\begin{eqnarray}
\begin{array}{l}
N_{8\mu}^N = \epsilon^{BCD} \lambda_{DA}^N (\tilde{q}_A \gamma_\nu q_B)
\Gamma^{\mu\nu}_{3/2} \gamma_5 q_C \, ,
\\ N_{9\mu}^N = \epsilon^{BCD} \lambda_{DA}^N (\tilde{q}_A \gamma_\nu \gamma_5 q_B)
\Gamma^{\mu\nu}_{3/2} q_C \, ,
\\ N_{10\mu}^N = \epsilon^{BCD} \lambda_{DA}^N (\tilde{q}_A \sigma_{\alpha\beta} q_B)
\Gamma^{\mu\alpha}_{3/2} \gamma^\beta \gamma_5 q_C \, .
\end{array}
\begin{array}{l}
N_{13\mu}^N = \epsilon^{CAD} \lambda_{DB}^N (\tilde{q}_A \gamma_\nu q_B)
\Gamma^{\mu\nu}_{3/2} \gamma_5 q_C \, ,\\
N_{14\mu}^N = \epsilon^{CAD} \lambda_{DB}^N (\tilde{q}_A \gamma_\nu \gamma_5 q_B)
\Gamma^{\mu\nu}_{3/2} q_C \, , \\
N_{15\mu}^N = \epsilon^{CAD} \lambda_{DB}^N (\tilde{q}_A \sigma_{\alpha\beta} q_B)
\Gamma^{\mu\alpha}_{3/2} \gamma^\beta \gamma_5 q_C \, .
\end{array}
\end{eqnarray}
Considering all three sets of bi-local fields, we find that they are related 
through the Fierz transformation:
\begin{eqnarray}
\nonumber N^N_{i\mu}(y,x,x) &\leftrightarrow& N^N_{i\mu}(x,y,x) \, ,
\\ \nonumber N^N_i(x,y,x) &=& T^{O2}_{ij} N^N_i(x,x,y) \, ,
\end{eqnarray}
where the flavor-octet Fierz matrix ${\bf T}^{O2}$ is
\begin{eqnarray}
{\bf T}^{O2} = \frac{1}{4} \left ( \begin{array}{ccc|ccc|ccc} 0 & 0 &
0 & -1 & 1 & 1 & 0 & 0 & 0
\\ 0 & 0 & 0 & -1 & 1 & -1 & 0 & 0 & 0
\\ 0 & 0 & 0 & -2 & -2 & 0 & 0 & 0 & 0
\\ \hline 0 & 0 &
0 & 0 & 0 & 0 & -1 & 1 & 1
\\ 0 & 0 & 0 & 0 & 0 & 0 & -1 & 1 & -1
\\ 0 & 0 & 0 & 0 & 0 & 0 & -2 & -2 & 0
\\ \hline  -1 & 1 & 1 & 0 & 0 &
0 & 0 & 0 & 0
\\ -1 & 1 & -1 & 0 & 0 & 0 & 0 & 0 & 0
\\ -2 & -2 & 0 & 0 & 0 & 0 & 0 & 0 & 0
\end{array} \right ) \, .
\end{eqnarray}
Together with the identity Eq. (\ref{eq:Jacobi}) and the Pauli principle
we find that two fields vanish identically:
\begin{eqnarray}
N^N_{4\mu}(x,x,y) = N^N_{5\mu}(x,x,y) = 0 \, .
\end{eqnarray}
Therefore, only three of the original 15 operators are independent.
Here we choose them as $N^N_{3\mu}(x,x,y)$ and
\begin{eqnarray}
M_{4\mu}(x,x,y) &=& N_{9\mu}(x,x,y) - N_{14\mu}(x,x,y) \, ,
\\ M_{5\mu}(x,x,y) &=& N_{10\mu}(x,x,y) - N_{15\mu}(x,x,y) \, .
\end{eqnarray}
Other bi-local octet baryons can be related to these three; here we only
show the representative equation for $N_{8\mu}(x,x,y)$:
\begin{eqnarray}
N_{8\mu}(x,x,y) &=& - \frac{1}{2} N_{3\mu}(x,x,y) \, ,
\end{eqnarray}

%
\subsection{Antisymmetric Tensor (Bargmann-Wigner) Fields}
\label{subsec:asT}
%

\subsubsection{Flavor singlet baryon}

We start by writing down the trilocal baryon field
\begin{eqnarray}
\begin{array}{l}
\Lambda_{5\mu\nu} = \epsilon^{ABC} (\tilde{q}_A \sigma_{\alpha\beta} q_B)
\Gamma^{\mu\nu\alpha\beta}_{3/2} q_C  \, ,
\end{array}
\end{eqnarray}
which contains a diquark formed wih the antisymmetric tensor matrices $\sigma_{\mu \nu}$.
Here $\Gamma^{\mu\nu\alpha\beta}$ is the Bargmann-Wigner projection operator
defined as
\begin{eqnarray}
\Gamma^{\mu\nu\alpha\beta}=\left(g^{\mu\alpha}g^{\nu\beta} -
\frac{1}{2} g^{\nu\beta}\gamma^\mu\gamma^\alpha +\half g^{\mu\beta}\gamma^\nu\gamma^\alpha +
\frac16 \sigma^{\mu\nu}\sigma^{\alpha\beta}\right)\, .
\end{eqnarray}
We have also three sets of bi-local fields that are related 
through the Fierz transformation:
\begin{eqnarray}
\nonumber \Lambda_{5\mu\nu}(y,x,x) &\leftrightarrow& \Lambda_{5\mu\nu}(x,y,x) \, , \\
\nonumber \Lambda_{5\mu}(x,y,x) &=& T^{S2}_{ij} \Lambda_{5\mu\nu}(x,x,y) \, ,
\end{eqnarray}
where the flavor-singlet Fierz (1x1 matrix) number ${\bf T}^{S3}$ is unity
\begin{eqnarray}
{\bf T}^{S3} = 1 \, .
\label{e:TS3}
\end{eqnarray}
The Pauli principle leads immediately to 
\begin{eqnarray}
\Lambda_{5\mu\nu}(x,x,y) = 0 \, .
\end{eqnarray}
Thus, we have obtained the result that all flavor-singlet bi-local
antisymmetric tensor fields vanish due to the Pauli principle.

\subsubsection{The flavour decuplet baryons}

Let us start with writing down the baryon field which contain
a diquark formed by the Dirac matrices $\sigma_{\mu \nu}$:
\begin{eqnarray}
\begin{array}{l}
\Delta^P_{5\mu\nu} = S_P^{ABC} (\tilde{q}_A \sigma_{\alpha\beta} q_B)
\Gamma^{\mu\nu\alpha\beta}_{3/2} q_C  \, .
\end{array}
\end{eqnarray}
We have also three sets of bi-local fields, and they are related to
each others through the Fierz transformation:
\begin{eqnarray}
\nonumber \Delta_{5\mu\nu}(y,x,x) &\leftrightarrow& \Delta_{5\mu\nu}(x,y,x) \, ,
\\ \nonumber \Delta_{5\mu\nu}(x,y,x) &=& T^{D3}_{ij} \Delta_{5\mu\nu}(x,x,y) \, ,
\end{eqnarray}
where the flavor-decuplet Fierz (1x1) matrix ${\bf T}^{D3}$ is equivalent to
the flavor-singlet Eq. (\ref{e:TS3})
\begin{eqnarray}
{\bf T}^{D3} = 1 \, .
\end{eqnarray}
Therefore, the only original operator is Pauli-allowed.

\subsubsection{The flavor octet baryon fields}
\label{subsub:octet_AST}

Start by writing down the flavor-octet trilocal baryon field
\begin{eqnarray}
\begin{array}{l}
N_{5\mu\nu}^N = \epsilon^{ABD} \lambda_{DC}^N (\tilde{q}_A
\sigma_{\alpha\beta} q_B) \Gamma^{\mu\nu\alpha\beta}_{3/2} q_C \, .
\end{array}
\end{eqnarray}
which contains a diquark formed by the $\sigma_{\mu \nu}$ matrices.
There are, however, also two other kinds of octet baryons with the flavor
structures $\epsilon^{BCD} \lambda_{DA}^N$ and $\epsilon^{CAD} \lambda_{DB}^N$:
\begin{eqnarray}
\begin{array}{l}
N_{10\mu\nu}^N = \epsilon^{BCD} \lambda_{DA}^N (\tilde{q}_A
\sigma_{\alpha\beta} q_B) \Gamma^{\mu\nu\alpha\beta}_{3/2} q_C \, .
\end{array}
\begin{array}{l}
N_{15\mu\nu}^N = \epsilon^{CAD} \lambda_{DB}^N (\tilde{q}_A
\sigma_{\alpha\beta} q_B) \Gamma^{\mu\nu\alpha\beta}_{3/2} q_C \, .
\end{array}
\end{eqnarray}
Considering all three sets of bi-local fields that are related 
through the Fierz transformation:
\begin{eqnarray}
\nonumber N^N_{i\mu\nu}(y,x,x) &\leftrightarrow& N^N_{i\mu\nu}(x,y,x) \, ,
\\ \nonumber N^N_{i\mu\nu}(x,y,x) &=& T^{O3}_{ij} N^N_{i\mu\nu}(x,x,y) \, ,
\end{eqnarray}
where the flavor-octet Fierz matrix ${\bf T}^{O3}$ is
\begin{eqnarray}
{\bf T}^{O3} = \left (
\begin{array}{ccc}
0 & 1 & 0
\\ 0 & 0 & 1
\\ 1 & 0 & 0
\end{array}
\right ) \, .
\end{eqnarray}
Together with the relation Eq. (\ref{eq:Jacobi}) and the Pauli principle
we find:
\begin{eqnarray}
N_{5\mu\nu}^N(x,x,y) = 0 \, .
\end{eqnarray}
Therefore, only one of the original three flavor-octet operators is
independent. Here we choose it as $M_{5\mu\nu}^N(x,x,y) =
N_{10\mu\nu}^N(x,x,y) - N_{15\mu\nu}^N(x,x,y)$.

\subsection{Summary of straight-forward bi-local fields}
\label{sub:summary}

We have investigated the chiral multiplets consisting of bi-local
three-quark baryon operators, where we took into account the Pauli
principle by way of the Fierz transformation. All spin $\half$ and
$\thalf$ baryon operators were classified according to their
Lorentz and isospin group representations, where spin and flavor
projection operators were employed in Table \ref{tab:summary}.
We have derived the non-trivial  Fierz relations among various baryon
operators and thus found the independent baryon fields, see Tables \ref{tab:spin12a},
\ref{tab:spin32a1}, and \ref{tab:spin32b1}.

Thus, for example in the spin $\frac12$ sector, three flavor
singlet fields (``$\Lambda$'s''), five octet fields (``nucleons''),
and two decimet fields (``$\Delta$'s'') are independent in the
bi-local limit, in stark contrast to the local limit where there
are (only) two nucleons and no $\Delta$, see Ref.
\cite{Nagata:2008zzc}. We see in Table \ref{tab:spin12a}, that
five out of 12 entries in the Table \ref{tab:summary} vanish in
the local operator limit $x \to y$, and other Fierz identities
reduce the number of independent chiral multiplets from seven to four.
\begin{table}[tbh]
\begin{center}
\caption{The Abelian and the non-Abelian axial charges (+ sign
indicates ``naive'', - sign ``mirror'' transformation properties)
and the non-Abelian chiral multiplets of spin $\frac12$, Lorentz
representation $(\frac{1}{2},0)$ nucleon $N$, delta resonance $\Delta$
and $\Lambda$ hyperon fields. All fields are independent and
Fierz invariant.
In the last column we show the
Fierz-equivalent/identical field in the local limit ($x \to y$).}
\begin{tabular}{ccccc}
\hline \hline
& $U_A(1)$ & $SU(3)_F$ & $SU_L(3) \times SU_R(3)$ & Fierz$(x\to y)_{\rm local ~lim.}$\\
\hline
$\Lambda_1 - \Lambda_2$ & $-1$ & $1$ & $(\mathbf{3},\bar{\mathbf{3}})\oplus(\bar{\mathbf{3}},\mathbf{3})$ & $\Lambda_3$ \\
$\Lambda_3$ & $-1$ & $1$  & $(\mathbf{3},\bar{\mathbf{3}})\oplus(\bar{\mathbf{3}},\mathbf{3})$ & $\Lambda_1 - \Lambda_2$ \\
$N_1 - N_2$ & $-1$ & $8$ & $(\mathbf{3},\bar{\mathbf{3}})\oplus(\bar{\mathbf{3}},\mathbf{3})$ & $N_3 - M_4$ \\
$N_3 - M_4$ & $-1$ & $8$  & $(\mathbf{3},\bar{\mathbf{3}})\oplus(\bar{\mathbf{3}},\mathbf{3})$ & $N_1 - N_2$ \\
$N_3 + \frac13 M_4$ & $-1$& $8$ & $(\mathbf{3},\mathbf{6})\oplus(\mathbf{6},\mathbf{3})$ & 0 \\
$\Delta_4$ & $-1$& $10$ & $(\mathbf{3},\mathbf{6})\oplus(\mathbf{6},\mathbf{3})$ & 0 \\
\hline
$\Lambda_1 + \Lambda_2$ & $+3$ & $1$ & $(\mathbf{1},\mathbf{1})$ & 0 \\
$N_1 + N_2$ & $+3$ & $8$ & $(\mathbf{8},\mathbf{1})\oplus(\mathbf{1},\mathbf{8})$ & $M_5$ \\
$M_5$ & $+3$& $8$ & $(\mathbf{8},\mathbf{1})\oplus(\mathbf{1},\mathbf{8})$ & $N_1 + N_2$ \\
$\Delta_5$ & $+3$& $10$ & $(\mathbf{10},\mathbf{1})\oplus(\mathbf{1},\mathbf{10})$ & 0 \\
\hline
\end{tabular}
\label{tab:spin12a}
\end{center}
\end{table}
\begin{table}[tbh]
\begin{center}
\caption{The Abelian and the non-Abelian axial charges and the
non-Abelian chiral multiplets of spin $\frac{3}{2}$, Lorentz
representation $(1,\frac{1}{2})$ nucleon and $\Delta$ fields. All
of the fields are independent and
Fierz-invariant. In the last column we show the
Fierz-equivalent/identical field in the local limit ($x \to y$).}
\begin{tabular}{ccccc}
\hline \hline
& $U_A(1)$ & $SU(3)_F$ & $SU_L(3) \times SU_R(3)$ & Fierz$(x\to y)_{\rm local ~lim.}$ \\
\hline
$\Lambda_3^{\mu}$ & $+1$ & $1$  & $(\bar{\mathbf{3}},\mathbf{3})\oplus(\mathbf{3},\bar{\mathbf{3}})$ & 0 \\
$N_3^{\mu} - M_4^{\mu}$ & $+1$ & $8$  & $(\bar{\mathbf{3}},\mathbf{3})\oplus(\mathbf{3},\bar{\mathbf{3}})$ & 0 \\
$N_3^{\mu} + \frac13 M_4^{\mu}$ & $+1$& $8$ & $(\mathbf{6},\mathbf{3})\oplus(\mathbf{3},\mathbf{6})$ & $N_5^{\mu}$ \\
$\Delta_4^{\mu}$ & $+1$& $10$ & $(\mathbf{6},\mathbf{3})\oplus(\mathbf{3},\mathbf{6})$ & $\Delta_5^{\mu}$ \\
$M_5^{\mu}$ & $+1$ & $8$ & $(\mathbf{6},\mathbf{3})\oplus(\mathbf{3},\mathbf{6})$ & $N_3^{\mu} + \frac13 N_4^{\mu}$ \\
$\Delta_5^{\mu}$ & $+1$ & $10$ & $(\mathbf{6},\mathbf{3})\oplus(\mathbf{3},\mathbf{6})$ & $\Delta_4^{\mu}$ \\
\hline
\end{tabular}
\label{tab:spin32a1}
\end{center}
\end{table}
The baryon fields $(\Lambda_1 - \Lambda_2, N_1 - N_2)$ and $(\Lambda_3, N_3 - M_4)$
form two independent $[(\mathbf{3},\bar{\mathbf{3}})\oplus(\bar{\mathbf{3}},\mathbf{3})]$
chiral multiplets; $N_1 + N_2$ and $M_5$ form two independent
$[(\mathbf{8},\mathbf{1})\oplus(\mathbf{1},\mathbf{8})]$
chiral multiplets; $( N_3 + \frac13 M_4, \Delta_4 )$ form one
$[(\mathbf{3},\mathbf{6})\oplus(\mathbf{6},\mathbf{3})]$ chiral multiplet;
$\Delta_5$ also forms a separate $[(\mathbf{10},\mathbf{1})\oplus(\mathbf{1},\mathbf{10})]$ chiral
multiplet.

In the spin $\frac32$ sector, the Rarita-Schwinger fields
$(\Lambda_{3}^{\mu}, N_{3}^{\mu} - M_{4}^{\mu})$ form an independent
$[(\bar{\mathbf{3}},\mathbf{3})\oplus(\mathbf{3},\bar{\mathbf{3}})]$ chiral multiplet
and $(N_{3}^{\mu} + \frac13 M_{4}^{\mu}, \Delta_{4}^{\mu} )$ and
$(M_{5}^{\mu},\Delta_{5}^{\mu})$ form two
$[(\mathbf{6},\mathbf{3})\oplus(\mathbf{3},\mathbf{6})]$ chiral multiplets, see
Table \ref{tab:spin32a1}. Similarly, Lorentz representation $(\frac{3}{2},0)$ Bargmann-Wigner fields
$M_{5}^{\mu\nu} \in [(\mathbf{8},\mathbf{1})\oplus(\mathbf{1},\mathbf{8})]$,
$\Delta_{5}^{\mu\nu} \in [(\mathbf{10},\mathbf{1})\oplus(\mathbf{1},\mathbf{10})]$
are also independent, see Table \ref{tab:spin32b1}. This is again in contrast with
the local limit where there is only one independent nucleon field and two independent
$\Delta$'s,~\cite{Nagata:2008zzc}.
\begin{table}[tbh]
\begin{center}
\caption{The Abelian and the non-Abelian axial charges and the
non-Abelian chiral multiplets of spin $\frac{3}{2}$, Lorentz
representation $(\frac{3}{2},0)$ nucleon and $\Delta$ fields. All
of the fields are independent and Fierz-invariant. In the last column we show the
Fierz-equivalent/identical field in the local limit ($x \to y$).}
\begin{tabular}{ccccc}
\hline \hline
& $U_A(1)$ & $SU(3)_F$ & $SU_L(3) \times SU_R(3)$ & Fierz$(x\to y)_{\rm local ~lim.}$ \\
\hline
$M_5^{\mu \nu}$ & $+3$& $8$ & $(\mathbf{8},\mathbf{1})\oplus(\mathbf{1},\mathbf{8})$ & 0 \\
$\Delta_5^{\mu \nu}$ & $+3$& $10$ & $(\mathbf{10},\mathbf{1})\oplus(\mathbf{1},\mathbf{10})$ & $\Delta_5^{\mu \nu}$ \\
\hline
\end{tabular}
\label{tab:spin32b1}
\end{center}
\end{table}

This exhausts all chiral multiplets obtained from straightforward
three-quark interpolators, so that relaxing the bi-local limit and
going to the tri-local case would not yield new chiral multiplets.
Note, however, that some chiral multiplets are repeated (doubled),
whereas their mirror image(s) do not appear: why? The answer to
this question has to do with the number (even/odd) of Dirac
$\gamma$-matrices that appear in the field itself. With one new
four-vector [the $(x-y)_{\mu}$] available, this problem is (very)
easily solved: contracting the spin $\frac32$ fields with this
four-vector yields new spin $\frac12$ fields.

\section{Non-Straightforward Three-Flavor Bi-local Three-Quark Fields}
\label{sec:non_straight_fields}
%

Thus far we have straightforwardly extended the local field
analysis to the bi-local case and thus ignored new, less
straightforward possibilities: besides the (CM variable $x$)
derivative $\partial_{\mu}$, we have one new four-vector [the
$(x-y)_{\mu}$] available. Contracting the various spin $\frac32$
fields with these four-vectors yields new spin $\frac12$ fields.

Once again we would like to note that the bilocal fields constructed in this
section may have components overlapping with more than one angular
momentum $J$ state.
Their chiral properties are independent of the exact value of $J$, however.

\subsection{Derivative-Contracted Fields}
\label{subsec:derivative1}

Contraction with the (CM variable $x$ in the local limit, or
$\frac13(2x + y)$ in the bi-local case) derivative
$\partial_{\mu}$ is obligatory, as the true Rarita-Schwinger fields
must satisfy the auxiliary condition $\partial^{\mu} \Psi_{\mu} =
0$, which is not automatically satisfied by Ioffe's three-quark
interpolators with one
Lorentz index $\mu$~\cite{Ioffe81,Espriu:1983hu}. Thus, one must
subtract the (generally non-vanishing) $\partial_{\mu}
\partial^{\nu} B_{\nu} \frac{1}{\partial_{\nu} \partial^{\nu}}$
from the original (un-subtracted) Ioffe fields $B_{\nu}$ in order
to obtain genuine Rarita-Schwinger fields
\begin{eqnarray}
\Psi_{\mu} = B_{\mu} - \partial_{\mu}\frac{\partial^{\nu}
B_{\nu}}{\partial_{\nu} \partial^{\nu}} \, .
\end{eqnarray}
That leaves us with $\Psi \simeq i \partial^{\nu} B_{\nu}$ as a new
Dirac field interpolator. One look at Table \ref{tab:nstrspin32a1} reveals
that these new fields have precisely the ``mirror'' properties to those
of the ``usual'', or ``naive" Dirac field interpolators in Table
\ref{tab:nstrspin12a}. Note, however, that the chiral multiplets
$[(\mathbf{1},\mathbf{1})]$, $[(\mathbf{8},\mathbf{1})\oplus(\mathbf{1},\mathbf{8})]$
and $[(\mathbf{10},\mathbf{1})\oplus(\mathbf{1},\mathbf{10})]$, and their
``mirror fields'' are still missing from this list of Rarita-Schwinger fields.

The same holds for Rarita-Schwinger fields obtained from the (local) Bargmann-Wigner
fields \cite{Chung:1981cc} by contraction with one derivative $\partial^{\nu}$:
\begin{eqnarray}
\Psi_{\mu} = \partial^{\nu} B_{\nu\mu} \, .
\end{eqnarray}
This takes care of the $[(\mathbf{8},\mathbf{1})\oplus(\mathbf{1},\mathbf{8})]$ and
$[(\mathbf{10},\mathbf{1})\oplus(\mathbf{1},\mathbf{10})]$
chiral multiplets, by way of Bargmann-Wigner fields
$\partial_{\nu} M_{5}^{\mu\nu} \in [(\mathbf{8},\mathbf{1})\oplus(\mathbf{1},\mathbf{8})]$,
$\partial_{\nu} \Delta_{5}^{\mu\nu} \in [(\mathbf{10},\mathbf{1})\oplus(\mathbf{1},\mathbf{10})]$,
but not of their mirror images, which are still missing from this
list of Rarita-Schwinger fields, as is the $[(\mathbf{1},\mathbf{1})]$ field.
Moreover, this procedure does not produce new Bargmann-Wigner fields with chiral
properties not seen thus far, see Table \ref{tab:nstrspin12a}.
We also note that one can not obtain new Dirac field interpolators from Bargmann-Wigner
fields due to the identity
$\partial^{\nu}\partial^{\mu} B_{\nu\mu} = \partial^{\mu}\partial^{\nu} B_{\nu\mu} = 0$.

In a short summary, the derivative-contracted fields produce new Dirac fields
$(\partial_{\mu}\Lambda_3^{\mu}, \partial_{\mu}(N_3^{\mu} - M_4^{\mu})) \in
[(\bar{\mathbf{3}},\mathbf{3})\oplus(\mathbf{3},\bar{\mathbf{3}})]$,
$(\partial_{\mu}(N_3^{\mu} + \frac13 M_4^{\mu}), \partial_{\mu}\Delta_4^{\mu}) \in
[(\mathbf{6},\mathbf{3})\oplus(\mathbf{3},\mathbf{6})]$ and
$(\partial_{\mu}M_5^{\mu}, \partial_{\mu}\Delta_5^{\mu}) \in
[(\mathbf{6},\mathbf{3})\oplus(\mathbf{3},\mathbf{6})]$,
and new Rarita-Schwinger fields
$\partial_{\nu} M_{5}^{\mu\nu} \in [(\mathbf{8},\mathbf{1})\oplus(\mathbf{1},\mathbf{8})]$ and
$\partial_{\nu} \Delta_{5}^{\mu\nu} \in [(\mathbf{10},\mathbf{1})\oplus(\mathbf{1},\mathbf{10})]$,
which do not vanish, see Tables \ref{tab:nstrspin12a} and Tables \ref{tab:nstrspin32a1}.

\subsection{Non-Derivative-Contracted Fields}
\label{subsec:non_derivative}

Similarly to previous subsection, we can contract Rarita-Schwinger
fields with the four-vector $(x-y)_{\mu}$ to
obtain new Dirac fields,
\begin{eqnarray}
\Psi = (x-y)^{\nu} B_{\nu}\, .
\end{eqnarray}
We can also contract Bargmann-Wigner fields with the four-vector
$(x-y)_{\mu}$ to obtain new Rarita-Schwinger fields,
\begin{eqnarray}
\Psi_{\mu} = (x-y)^{\nu} B_{\nu\mu} \, .
\end{eqnarray}
Again we can not obtain new Dirac field interpolators from Bargmann-Wigner
fields due to $(x-y)^{\nu}(x-y)^{\mu} B_{\nu\mu} = (x-y)^{\mu}(x-y)^{\nu} B_{\nu\mu} = 0$.

In a short summary, the non-derivative-contracted fields produce new Dirac fields
$((x-y)_{\mu}\Lambda_3^{\mu}, (x-y)_{\mu}(N_3^{\mu} - M_4^{\mu}))$,
$((x-y)_{\mu}(N_3^{\mu} + \frac13 M_4^{\mu}), (x-y)_{\mu}\Delta_4^{\mu})$
and $((x-y)_{\mu}M_5^{\mu}, (x-y)_{\mu}\Delta_5^{\mu})$, and new Rarita-Schwinger fields
$(x-y)_{\nu} M_{5}^{\mu\nu}$ and $(x-y)_{\nu} \Delta_{5}^{\mu\nu}$,
see Tables \ref{tab:nstrspin12a} and \ref{tab:nstrspin32a1}. The chiral representations of these fields are the same as the
corresponding derivative-contracted fields.
Fierz identities show that all of these fields vanish in the local limit $x
\to y$.

\begin{table}[tbh]
\begin{center}
\caption{The Abelian and the non-Abelian axial charges
and the non-Abelian chiral multiplets of spin $\frac12$, Lorentz
representation $(\frac{1}{2},0)$, non-straightforward ``nucleon'' $N$ octet, delta resonance $\Delta$ decuplet and $\Lambda$ hyperon
singlet fields. All fields are independent and Fierz invariant.}
\begin{tabular}{ccccc}
\hline \hline
& $U_A(1)$ & $SU(3)_F$ & $SU_L(3) \times SU_R(3)$ & Fierz$(x\to y)_{\rm local ~lim.}$\\
\hline
$\partial_{\mu}\Lambda_3^{\mu}$ & $+1$ & $1$  & $(\bar{\mathbf{3}},\mathbf{3})\oplus(\mathbf{3},\bar{\mathbf{3}})$ & 0 \\
$\partial_{\mu}(N_3^{\mu} - M_4^{\mu})$ & $+1$ & $8$  & $(\bar{\mathbf{3}},\mathbf{3})\oplus(\mathbf{3},\bar{\mathbf{3}})$ & 0 \\
$\partial_{\mu}(N_3^{\mu} + \frac13 M_4^{\mu})$ & $+1$& $8$ & $(\mathbf{6},\mathbf{3})\oplus(\mathbf{3},\mathbf{6})$ & $\partial_{\mu}M_5^{\mu}$ \\
$\partial_{\mu}\Delta_4^{\mu}$ & $+1$& $10$ & $(\mathbf{6},\mathbf{3})\oplus(\mathbf{3},\mathbf{6})$ & $\partial_{\mu}\Delta_5^{\mu}$ \\
$\partial_{\mu}M_5^{\mu}$ & $+1$ & $8$ & $(\mathbf{6},\mathbf{3})\oplus(\mathbf{3},\mathbf{6})$ & $\partial_{\mu}(N_3^{\mu} + \frac13 M_4^{\mu})$ \\
$\partial_{\mu}\Delta_5^{\mu}$ & $+1$ & $10$ & $(\mathbf{6},\mathbf{3})\oplus(\mathbf{3},\mathbf{6})$ & $\partial_{\mu}\Delta_4^{\mu}$ \\
\hline
$(x-y)_{\mu}\Lambda_3^{\mu}$ & $+1$ & $1$  & $(\bar{\mathbf{3}},\mathbf{3})\oplus(\mathbf{3},\bar{\mathbf{3}})$ & 0 \\
$(x-y)_{\mu}(N_3^{\mu} - M_4^{\mu})$ & $+1$ & $8$  & $(\bar{\mathbf{3}},\mathbf{3})\oplus(\mathbf{3},\bar{\mathbf{3}})$ & 0 \\
$(x-y)_{\mu}(N_3^{\mu} + \frac13 M_4^{\mu})$ & $+1$& $8$ & $(\mathbf{6},\mathbf{3})\oplus(\mathbf{3},\mathbf{6})$ & 0 \\
$(x-y)_{\mu}\Delta_4^{\mu}$ & $+1$& $10$ & $(\mathbf{6},\mathbf{3})\oplus(\mathbf{3},\mathbf{6})$ & 0 \\
$(x-y)_{\mu}M_5^{\mu}$ & $+1$ & $8$ & $(\mathbf{6},\mathbf{3})\oplus(\mathbf{3},\mathbf{6})$ & 0 \\
$(x-y)_{\mu}\Delta_5^{\mu}$ & $+1$ & $10$ & $(\mathbf{6},\mathbf{3})\oplus(\mathbf{3},\mathbf{6})$ & 0 \\
\hline
$(x-y)_\mu\partial_{\nu}M_5^{\mu \nu}$ & $+3$& $8$ & $(\mathbf{8},\mathbf{1})\oplus(\mathbf{1},\mathbf{8})$ & 0 \\
$(x-y)_\mu\partial_{\nu}\Delta_5^{\mu \nu}$ & $+3$& $10$ & $(\mathbf{10},\mathbf{1})\oplus(\mathbf{1},\mathbf{10})$ & 0 \\
\hline
\end{tabular}
\label{tab:nstrspin12a}
\end{center}
\end{table}

\subsection{Mixed-Contracted Fields}
\label{subsec:derivative2}

Together with the derivative and the four-vector $(x-y)_{\mu}$,
we obtain new Dirac fields from Bargmann-Wigner fields
\begin{eqnarray}
\Psi = (x-y)^{\nu} \partial^{\mu} B_{\nu\mu} \, .
\end{eqnarray}
The other three $(x-y)^{\mu} \partial^{\nu} B_{\nu\mu}$,
$\partial^{\nu} (x-y)^{\mu} B_{\nu\mu}$, and
$\partial^{\mu} (x-y)^{\nu} B_{\nu\mu}$ can be related to this one,
and so this is the only independent field.
Therefore, the mixed-contracted fields only produce the Dirac fields
$(x-y)_\mu\partial_{\nu}M_5^{\mu \nu} \in [(\mathbf{8},\mathbf{1})\oplus(\mathbf{1},\mathbf{8})]$
and $(x-y)_\mu\partial_{\nu}\Delta_5^{\mu \nu} \in [(\mathbf{10},\mathbf{1})\oplus(\mathbf{1},\mathbf{10})]$,
all of which vanish in the local limit $x \to y$, see Table \ref{tab:nstrspin12a}.

\begin{table}[tbh]
\begin{center}
\caption{The Abelian and the non-Abelian axial charges and the
non-Abelian chiral multiplets of spin $\frac{3}{2}$, Lorentz
representation $(1,\frac{1}{2})$ ``nucleon'' octet and ``$\Delta$''
decuplet non-straightforward fields.}
\begin{tabular}{ccccc}
\hline \hline
& $U_A(1)$ & $SU(3)_F$ & $SU_L(3) \times SU_R(3)$ & Fierz$(x\to y)_{\rm local ~lim.}$ \\
\hline
$\partial_{\nu} M_{5}^{\mu\nu}$ & $+3$ & $8$ & $(\mathbf{8},\mathbf{1})\oplus(\mathbf{1},\mathbf{8})$ & 0 \\
$\partial_{\nu} \Delta_{5}^{\mu\nu}$ & $+3$ & $10$ & $(\mathbf{10},\mathbf{1})\oplus(\mathbf{1},\mathbf{10})$
& $\partial_{\nu} \Delta_{5}^{\mu\nu}$ \\
\hline
$(x-y)_{\nu} M_{5}^{\mu\nu}$ & $+3$ & $8$ & $(\mathbf{8},\mathbf{1})\oplus(\mathbf{1},\mathbf{8})$ & 0 \\
$(x-y)_{\nu} \Delta_{5}^{\mu\nu}$ & $+3$ & $10$ & $(\mathbf{10},\mathbf{1})\oplus(\mathbf{1},\mathbf{10})$ & 0 \\
\hline
\end{tabular}
\label{tab:nstrspin32a1}
\end{center}
\end{table}

\section{Summary and Conclusions}
\label{sec:summary}

We have investigated the chiral multiplets consisting of bi-local
three-quark baryon operators, where we took into account the Pauli
principle by way of the Fierz transformation. All spin $\frac12$ and
some $\frac32$ baryon operators were classified in Tables \ref{tab:spin12a},
\ref{tab:spin32a1}, \ref{tab:spin32b1}, \ref{tab:nstrspin12a} and \ref{tab:nstrspin32a1},
according to their Lorentz and flavor symmetry group representations. Again we would like to note that
these baryon fields have definite total angular momentum only in the local limit.
We have employed the standard flavor $SU(3)$ formalism instead of the
explicit expressions in terms of different flavored quarks in the
flavor components of the baryon fields that are commonplace in
this line of work.

In doing so, we have been able to systematically
derive the Fierz identities and chiral transformations of the baryon fields.
More specifically, we have derived all non-trivial Fierz relations
among various baryon bi-local operators and thus found the independent bi-local
baryon fields. We have shown that the Fierz transformation connects only those bi-local
baryon interpolating fields with identical chiral group-theoretical properties, i.e.,
those belonging to the same chiral multiplet, just as in the case of local baryon
operators.

For example, in the spin $\frac12$ sector, five flavor singlet fields
(``$\Lambda$'s''), 12 octet fields (``nucleons''),
and seven decimet fields (``$\Delta$'s'') are independent in the
bi-local limit, in stark contrast to the local limit where there
is (only) one $\Lambda$, two nucleons and no $\Delta$'s, Ref.~\cite{Nagata:2008zzc}.
One can see that 14 out of 24 entries in the Tables \ref{tab:spin12a}
and \ref{tab:nstrspin12a} vanish in the local operator limit $x
\to y$, and another three Fierz identities reduce the number of
independent fields from 10 to five.

The $(\Lambda_1 + \Lambda_2)$ forms one independent $[(\mathbf{1},{\mathbf{1}})]$
chiral multiplet, $\left(\Lambda_1 - \Lambda_2, N_1 - N_2 \right)$ and
$(\Lambda_3, {N_3} - M_4)$ form two independent
$[(\mathbf{3},\bar{\mathbf{3}})\oplus(\bar{\mathbf{3}},\mathbf{3})]$ chiral multiplets,
$\left(N_1 + N_2 \right)$ and $M_5$ form two independent
$[(\mathbf{8},\mathbf{1})\oplus(\mathbf{1},\mathbf{8})]$ chiral multiplets,
$\left({N_3} + \frac13 M_4, \Delta_4
\right)$ form one $[(\mathbf{3},\mathbf{6})\oplus(\mathbf{6},\mathbf{3})]$
chiral multiplet and the independent field $\Delta_5$ also forms a separate
$[(\mathbf{10},\mathbf{1})\oplus(\mathbf{1},\mathbf{10})]$ chiral multiplet.

The derivative-contracted fields produce new non-vanishing Dirac fields
$(\partial_{\mu}\Lambda_3^{\mu}, \partial_{\mu}(N_3^{\mu} - M_4^{\mu})) \in
[(\bar{\mathbf{3}},\mathbf{3})\oplus(\mathbf{3},\bar{\mathbf{3}})]$,
$(\partial_{\mu}(N_3^{\mu} + \frac13 M_4^{\mu}), \partial_{\mu}\Delta_4^{\mu}) \in
[(\mathbf{6},\mathbf{3})\oplus(\mathbf{3},\mathbf{6})]$ and
$(\partial_{\mu}M_5^{\mu}, \partial_{\mu}\Delta_5^{\mu}) \in
[(\mathbf{6},\mathbf{3})\oplus(\mathbf{3},\mathbf{6})]$.
The non-derivative-contracted fields produce
$((x-y)_{\mu}\Lambda_3^{\mu}, (x-y)_{\mu}(N_3^{\mu} - M_4^{\mu}))$,
$((x-y)_{\mu}(N_3^{\mu} + \frac13 M_4^{\mu}), (x-y)_{\mu}\Delta_4^{\mu})$
and $((x-y)_{\mu}M_5^{\mu}, (x-y)_{\mu}\Delta_5^{\mu})$, and
the mixed-contracted fields produce
$(x-y)_\mu\partial_{\nu}M_5^{\mu \nu} \in [(\mathbf{8},\mathbf{1})\oplus(\mathbf{1},\mathbf{8})]$
and $(x-y)_\mu\partial_{\nu}\Delta_5^{\mu \nu} \in [(\mathbf{10},\mathbf{1})\oplus(\mathbf{1},\mathbf{10})]$,
see Tables \ref{tab:nstrspin12a}.

In the spin $\frac32$ sector, the $(\Lambda_{3}^{\mu}, N_{3}^{\mu} - M_{4}^{\mu})$ form an
independent $[(\mathbf{3},\bar{\mathbf{3}})\oplus(\bar{\mathbf{3}},\mathbf{3})]$ chiral multiplet,
whereas
$\left(N_{3}^{\mu} + \frac13 M_{4}^{\mu}, \Delta_{4}^{\mu} \right) \in
[(\mathbf{6},\mathbf{3}) \oplus (\mathbf{3},\mathbf{6})]$
and
$(M_{5}^{\mu},\Delta_{5}^{\mu}) \in [(\mathbf{6},\mathbf{3}) \oplus (\mathbf{3},\mathbf{6})]$,
are also independent, again in contrast with the local limit where there is only one independent
nucleon field and two independent $\Delta$'s,~\cite{Nagata:2008zzc}.
The derivative-contracted fields produce new Rarita-Schwinger fields
$\partial_{\nu} M_{5}^{\mu\nu} \in [(\mathbf{8},\mathbf{1})\oplus(\mathbf{1},\mathbf{8})]$ and
$\partial_{\nu} \Delta_{5}^{\mu\nu} \in [(\mathbf{10},\mathbf{1})\oplus(\mathbf{1},\mathbf{10})]$.
The non-derivative-contracted fields produce $(x-y)_{\nu} M_{5}^{\mu\nu}$
and $(x-y)_{\nu} \Delta_{5}^{\mu\nu}$, see Tables \ref{tab:nstrspin32a1}.

This increase of the number of independent fields is in line with
expectations based on the non-relativistic quark model, where the
number of Pauli-allowed three-quark states in the $L^P = 1^-$
shell sharply rises from the corresponding number in the ground
state. Indeed, there is a deep analogy between the Pauli principle
acting in the non-relativistic quantum formalism, where the flavor-spin
group $SU(6)_{FS}$ plays the role of the chiral symmetry group
$SU_L(3)\times SU_R(3)$ in the relativistic formalism.
Of course, the chiral symmetry group $SU_L(3)\times SU_R(3)$ is a subgroup
of some (``bigger'') $SU(6)$, 
but that is not the flavor-spin group $SU(6)_{FS}$\footnote{There is, of course,
the late-1960's and early-1970's work, Refs.
\cite{Gilman:1967qs,Gilman:1972dm,Gilman:1973kh,Melosh:1974cu,Close:1974ux,Carlitz:1976in}
about the $SU_{W}(6)$ and the ``collinear'' chiral symmetry group
$SU_L(3)\times SU_R(3) \times O(2)$, which is not even invariant
under all spatial rotations, let alone under (all) Lorentz transformations.}.
This analogy is at the present still (only) empirical: we do not have a set of clear
and simple rules that determine the allowed chiral multiplets in this relativistic
approach, that would correspond to the rules leading to the allowed $SU_{FS}(6)$ multiplets
in the non-relativistic approach. Rather, we had to rely on the (rather involved) present
analysis.

The physical significance of our present work is that it shows that there is no need to
introduce $q {\bar q}$ components in addition to the three-quark ``core'', so as to agree
with the observed axial couplings and magnetic moments: the phenomenologically necessary
$[(\mathbf{6},\mathbf{3}) \oplus (\mathbf{3},\mathbf{6})]$ chiral component and the 
$[(\mathbf{3},\bar{\mathbf{3}})\oplus(\bar{\mathbf{3}},\mathbf{3})]$
``mirror'' component exist as bi-local fields \footnote{This is not to say that all
chiral multiplets exist in the three-quark limit: the ``mirror'' chiral multiplets
$[(\mathbf{1},\mathbf{10})\oplus(\mathbf{10},\mathbf{1})]$ and
$[(\mathbf{1},\mathbf{8})\oplus(\mathbf{8},\mathbf{1})]$ do not show up in our tables.}.
Thus, we have shown that there is no need for ``meson cloud'', or (non-exotic) ``pentaquark''
components in the Fock expansion of the baryon wave function, to explain (at least)
the axial currents and magnetic moments, contrary to established opinion,
Ref. \cite{Donoghue:1985bu}. This goes to show that the algebraic complexity
of three Dirac quark fields is such that it can mimick the presence of $q {\bar q}$
pairs, at least in certain observables. For us this was a surprise.

The framework presented here holds in standard approaches to QCD,
such as the QCD sum rules \cite{Ioffe81,Espriu:1983hu} and lattice QCD
\cite{Basak:2005ir}, 
under the proviso that chiral symmetry is observed by the approximation used.
There is another (sub-)field of QCD where it ought to make an impact: in the class
of fully relativistic three-body models, such as those based on the three-body Salpeter,
Refs.~\cite{Henriques:1975uh,Loring:2001kv,Kvinikhidze:2007qq}, or Bethe-Salpeter equation
approaches to chiral quark models Refs.~\cite{King:1986wi,Sotiropoulos:1994ub,Stefanis:1999wy,Eichmann:2009qa}.
One potential application of our results is to classify
various components in the Salpeter, or Bethe-Salpeter amplitudes (wave functions),
instead of the non-relativistic $SU(6)_{FS}$ multiplets that have been used so far,
and thus to try and determine the baryons' chiral mixing coefficients (angles),
Refs.~\cite{Dmitrasinovic:2009vp,Dmitrasinovic:2009vy,Chen:2009sf,Chen:2010ba},
starting from an underlying chiral model. Model calculations like that could give one
insight into structural questions that cannot be (reasonably) expected to be answered
by lattice QCD. For example, why do certain chiral multiplets not appear in the
baryons?

\section*{Acknowledgments}

The work of H.X.C is supported by the National Natural Science Foundation of China under
Grant No. 11205011, and the Fundamental Research Funds for the Central Universities. The work of V.D. was supported by the Serbian
Ministry of Science and Technological Development under grant numbers OI 171037 and III 41011.

\appendix

\section{Chiral Transformations}
\label{sec:chiral_baryon}

Here, we briefly review the $SU_L(3)\times SU_R(3)$ chiral transformations of
three-quark baryon operators, which are determined by their Dirac matrix structure, see
Ref. \cite{Chen:2008qv}.
Under the $U_V(1)=U_B(1)$ (baryon number), $U_A(1)$ (axial baryon number),
$SU_V(3)=SU_F(3)$ (flavor SU(3)) and $SU_A(3)$ (axial flavor SU(3)) transformations, the
quark field, $q= q_L + q_R$, transforms as
\begin{eqnarray}
\nonumber
\bf{U_{V}(1)} &:& q \to \exp(i a^0) q  = q + \delta q \, ,
\\
\bf{SU_V(3)} &:& q \to \exp (i \vec \lambda \cdot \vec a ){q} = q + \delta^{\vec{a}} q \, ,
\\ \nonumber
\bf{U_{A}(1)} &:& q \to \exp(i \gamma_5 b^0) q = q + \delta_5 q \, ,
\\ \nonumber
\bf{SU_{A}(3)} &:& q \to \exp (i \gamma_{5} \vec \lambda \cdot \vec b){q} = q + \delta_5^{\vec{b}} q \, ,
\end{eqnarray}
where $\vec \lambda$ are the eight Gell-Mann matrices; $a^0$ is the infinitesimal
parameter for the $U_V(1)$ ``vector'' transformation, $\vec{a}$ are the octet of $SU_V(3)$ group
parameters, $b^0$ is the infinitesimal parameter for the $U_A(1)$ $\gamma_5$ transformation,
and $\vec{b}$ are the octet of $SU_A(3)$ $\gamma_5$ transformation parameters.

The $U_V(1)$ baryon number (``vector'') transformation is simple, while the $SU_V(3)$
flavor-symmetry (``vector'') transformations are also well known:
\begin{enumerate}

\item for any singlet baryon field $\Lambda$ , we have
\begin{eqnarray}
\delta^{\vec a} \Lambda &=& 0 \, ;
\end{eqnarray}

\item for any octet baryon field $N^M$, we have
\begin{eqnarray}
\delta^{\vec a} N^M &=& 2 a^N f_{NMO} N^O \, ;
\end{eqnarray}

\item for any decuplet baryon field $\Delta^P$, we have
\begin{eqnarray}
\delta^{\vec a} \Delta^P &=& 2 i a^N {\bf F}^N_{PQ} \Delta^Q \, ,
\end{eqnarray}

\end{enumerate}
where the coefficients $d^{NMO}$ and $f^{NMO}$ are the standard symmetric and antisymmetric
structure constants of $SU(3)$; the transition matrices ${\bf F}^N_{PQ}$ as well as
${\bf T}^{\dagger N}_{PM}$ in the following subsections are listed in Ref.~\cite{Chen:2009sf}.

\subsection{Dirac fields (spin $\frac{1}{2}$)}

Under the Abelian chiral transformation the rule, we have
\begin{eqnarray}
\delta_5 ( \Lambda_1 + \Lambda_2 ) &=& 3 i b \gamma_5 ( \Lambda_1
+ \Lambda_2 ) \, ,
\\ \delta_5 ( \Lambda_1 - \Lambda_2 ) &=& - i b \gamma_5 ( \Lambda_1 - \Lambda_2 ) \, ,
\\ \delta_5 \Lambda_3 &=& - i b \gamma_5 \Lambda_3 \, ,
\end{eqnarray}
and
\begin{eqnarray}
\delta_5 \Delta^P_4 &=& - i b \gamma_5 \Delta^P_4 \, ,
\\ \delta_5 \Delta^P_5 &=& 3 i b \gamma_5 \Delta^P_5 \, ,
\end{eqnarray}
and
\begin{eqnarray}
\delta_5 ( N^N_1 + N^N_2 ) &=& 3 i b \gamma_5 ( N^N_1 + N^N_2 ) \,
,
\\ \delta_5 ( N^N_1 - N^N_2 ) &=& - i b \gamma_5 ( N^N_1 - N^N_2 ) \, ,
\\ \delta_5 N^N_3 &=& - i b \gamma_5 N^N_3 \, ,
\\ \delta_5 M^N_4 &=& - i b \gamma_5 M^N_4 \, ,
\\ \delta_5 M^N_5 &=& 3 i b \gamma_5 M^N_5 \, .
\end{eqnarray}

Under the $SU_A(3)$ chiral transformation the rule, we have
\begin{eqnarray}
\delta_5^{\vec{b}} ( \Lambda_1 + \Lambda_2 ) &=& 0 \, ,
\\ \delta_5^{\vec{b}} ( \Lambda_1 - \Lambda_2 ) &=& 2 i b^N \gamma_5 ( N^N_1 - N^N_2 ) \, ,
\\ \delta_5^{\vec{b}} \Lambda_3 &=& - i b^N \gamma_5 ( N^N_3 - M_4^N ) \, ,
\end{eqnarray}
and
\begin{eqnarray}
\delta_5^{\vec{b}} \Delta_4^P &=& i b^N \gamma_5 {\bf T}^{\dagger
N}_{PM} ( N^M_3 + \frac{1}{3} M^M_4 ) - \frac{2}{3} i b^N \gamma_5
{\bf F}^N_{PQ} \Delta^Q_4 \, ,
\\ \delta_5^{\vec{b}} \Delta_5^P &=& 2 i b^N \gamma_5 {\bf F}^N_{PQ} \Delta^Q_5 \, ,
\end{eqnarray}
and
\begin{eqnarray}
\delta_5^{\vec{b}} ( N^M_1 + N^M_2 ) &=& 2 b^N \gamma_5 f^{NMO} (
N^O_1 + N^O_2 ) \, ,
\\ \delta_5^{\vec{b}} ( N^M_1 - N^M_2 ) &=& \frac{4}{3} i b^N \gamma_5 ( \Lambda_1 - \Lambda_2 )
+ 2 i b^N \gamma_5 d^{NMO} ( N^O_1 - N^O_2 ) \, ,
\\ \delta_5^{\vec{b}} ( N^M_3 +\frac{1}{3} M^M_4 ) &=& \frac{16}{3} i b^N \gamma_5
{\bf T}^N_{MP} \Delta^P_4 + i b^N \gamma_5 ( - 2 d^{NMO} + \frac{4}{3} i f^{NMO})
( N^O_3 + \frac{1}{3} M^O_4 ) \, ,
\\ \delta_5^{\vec{b}} ( N^M_3 - M^M_4 ) &=& - \frac{8}{3} i b^N \gamma_5 \Lambda_3 +
2 i b^N \gamma_5 d^{NMO} ( N^O_3 - M^O_4 ) \, ,
\\ \delta_5^{\vec{b}} M^M_5 &=& 2 b^N \gamma_5 f^{NMO} M^O_5 \, .
\end{eqnarray}

\subsection{Rarita-Schwinger fields  (spin $\frac{1}{2}$ and $\frac{3}{2}$)}

Under the Abelian chiral transformation, we have
\begin{eqnarray}
\delta_5 \Lambda_{3\mu} &=& i b \gamma_5 \Lambda_{3\mu} \, ,
\end{eqnarray}
and
\begin{eqnarray}
\delta_5 \Delta^P_{4\mu} &=& i b \gamma_5 \Delta^P_{4\mu} \, ,
\\ \delta_5 \Delta^P_{5\mu} &=& i b \gamma_5 \Delta^P_{5\mu} \, ,
\end{eqnarray}
and
\begin{eqnarray}
\delta_5 N^N_{3\mu} &=& i b \gamma_5 N^N_{3\mu} \, ,
\\ \delta_5 M^N_{4\mu} &=& i b \gamma_5 M^N_{4\mu} \, ,
\\ \delta_5 M^N_{5\mu} &=& i b \gamma_5 M^N_{5\mu} \, .
\end{eqnarray}

Under the $SU_A(3)$ chiral transformations, we have
\begin{eqnarray}
\delta_5^{\vec{b}} \Lambda_{3\mu} &=& i b^N \gamma_5 ( N^N_{3\mu}
- M_{4\mu}^N ) \, ,
\end{eqnarray}
and
\begin{eqnarray}
\delta_5^{\vec{b}} \Delta_{4\mu}^P &=& - i b^N \gamma_5 {\bf
T}^{\dagger N}_{PM} ( N^M_{3\mu} + \frac{1}{3} M^M_{4\mu} ) +
\frac{2}{3} i b^N \gamma_5 {\bf F}^N_{PQ} \Delta^Q_{4\mu} \, ,
\\ \delta_5^{\vec{b}} \Delta_{5\mu}^P &=& \frac{2}{3} i b^N \gamma_5
{\bf T}^{\dagger N}_{PM} M^M_{5\mu} + \frac{2}{3} i b^N \gamma_5
{\bf F}^N_{PQ} \Delta^Q_{5\mu} \, ,
\end{eqnarray}
and
\begin{eqnarray}
\delta_5^{\vec{b}} ( N^M_{3\mu} + \frac{1}{3} M^M_{4\mu} ) &=& -
\frac{16}{3} i b^N \gamma_5 {\bf T}^N_{MP} \Delta^P_{4\mu} + i b^N
\gamma_5 ( 2 d^{NMO} - \frac{4}{3} i f^{NMO} ) ( N^O_{3\mu} +
\frac{1}{3} M^O_{4\mu} ) \, ,
\\ \delta_5^{\vec{b}} ( N^M_{3\mu} - M^M_{4\mu} ) &=& \frac{8}{3} i b^N \gamma_5 \Lambda_{3\mu}
- 2 i b^N \gamma_5 d^{NMO} ( N^O_{3\mu} - M^O_{4\mu} ) \, ,\\
\delta_5^{\vec{b}} M^M_{5\mu} &=& 8 i b^N \gamma_5 {\bf T}^N_{MP} \Delta^P_{5\mu} +
i b^N \gamma_5 ( 2 d^{NMO} - \frac{4}{3} i f^{NMO} ) ( N^O_{3\mu} + \frac{1}{3} M^O_{4\mu} ) \, .
\end{eqnarray}

\subsection{Antisymmetric Tensor fields (spin $\frac{3}{2}$)}

Under the Abelian chiral transformation, we have
\begin{eqnarray}
\delta_5 \Delta^P_{5\mu\nu} &=& 3 i b \gamma_5 \Delta^P_{5\mu\nu}
\, ,
\end{eqnarray}
and
\begin{eqnarray}
\delta_5 M^N_{5\mu\nu} &=& 3 i b \gamma_5 M^N_{5\mu\nu} \, .
\end{eqnarray}

Under the $SU_A(3)$ chiral transformations, we have
\begin{eqnarray}
\delta_5^{\vec{b}} \Delta_{5\mu\nu}^P &=& i b^N \gamma_5 {\bf
F}^N_{PQ} \Delta^Q_{5\mu\nu} \, ,
\end{eqnarray}
and
\begin{eqnarray}
\delta_5^{\vec{b}} M^M_{5\mu\nu} &=& 2 b^N \gamma_5 f^{NMO}
M^O_{5\mu\nu} \, .
\end{eqnarray}

\end{document}